\documentclass[12pt,letterpaper]{article}
\usepackage{amsmath,amssymb,array,calc,rotating,epsfig,psfrag, amscd, datetime}
\usepackage{color}
\usepackage[left=2cm,top=2cm,right=2cm,nohead]{geometry}

\def\bc{\begin{center}}
\def\ec{\end{center}}
\def\bea{\begin{eqnarray}}
\def\eea{\end{eqnarray}}
\def\beq{\begin{equation}}
\def\eeq{\end{equation}}
\def\nn{\nonumber}
\def\dd{{\rm d}}

\def\del{\partial}

\def \mt   {\ifmmode m_{\rm t} \else $m_{\rm t}$ \fi}

\newcommand     \MSB            {\ifmmode {\overline{\rm MS}} \else
                                 $\overline{\rm MS}$  \fi}

\def\sla#1{\rlap{\begin{picture}(10,10)
\put(0,0){\line(1,1){10}}
\end{picture} }#1}
\def\slash#1#2#3{\rlap{\begin{picture}(10,10)
\put(0,0){\line(#1,1){#2}}
\end{picture}}#3}

\begin{document}

\begin{center}
\vskip 2 cm

{\Large \bf  A Note on  Supersymmetric \\ \vskip 0.25cm Type II Solutions of Lifshitz Type}
\vskip 1.25 cm 
{Michela Petrini$^{a,b}$ and Alberto Zaffaroni$^{c}$}\\
\vskip 10mm

$^a$Laboratoire de Physique Th\' eorique et
Hautes Energies, \\   Universit\' e
Pierre et Marie Curie, \\ 
4 Place Jussieu, 75252 Paris Cedex 05,
France \\
\vskip 5mm
$^b$Institut de Physique Th\'eorique, \\
CEA Saclay, CNRS URA 2306, \\
F-91191 Gif-sur-Yvette, France \\
\vskip 5mm
$^c$Dipartimento di Fisica, Universit\`a di Milano--Bicocca, I-20126 Milano, Italy\\
            and\\
            INFN, sezione di Milano--Bicocca,
            I-20126 Milano, Italy
\end{center}
\vskip 2cm
\begin{abstract}
We discuss a  class of supersymmetric type II non-relativistic solutions with exact or asymptotic scale invariance.
As already  emerged from previous investigations, we find a clear  correspondence between anisotropic $d$-dimensional vacua and relativistic solutions in $d+1$ dimensions.
We will show that  supersymmetric four-dimensional Poincar\'e invariant backgrounds  in type IIB can descend to analogous solutions with 
anisotropic scaling in $t$ and $(x,y)$. This result can be applied to scale invariant theories,   domain walls interpolating between four-dimensional 
Lifshitz vacua  and more general solutions with only asymptotic, approximate scaling behaviour and hyperscaling violation.
\end{abstract}

\newpage
\section{Introduction}

The gauge/gravity correspondence relates  relativistic strongly coupled gauge theories to supergravity on Anti de Sitter or warped Minkowski space-times.
Recently, the application of the ideas of holography to condensed matter systems have motivated the study of non relativistic geometries in 
string/supergravity theories.  For instance, systems of strongly correlated electrons can exhibit critical points with an anisotropic rescaling between time and space
\beq
t \rightarrow \lambda^{ z} t \qquad \qquad x^i \rightarrow \lambda x^i \quad  i = 1 \ldots D \, .
\eeq
According to the holographic dictionary such behaviour should be described by a Lifshitz geometry
\beq
\dd s^2 = -  r^{2 z} \dd t^2 + r^2\, \sum_{i=1}^D  (\dd x^i)^2 + \frac{\dd r^2}{r^2} \, , 
\eeq
where $r$ is the holographic energy direction. 

Contrary to other non-relativistic solutions, like Schr\"odinger geometries, embedding Lifshitz solutions in string theory turned out to be a relatively non trivial issue. Lifshitz geometries  appear in  four dimensional models of gravity coupled to a topological term or a massive vector \cite{KLM08, T08}.  While initially some no-go theorems seemed to exclude the possibility of embedding Lifshitz solutions in the full supergravity theory\footnote{ See for example \cite{LNT09, BDV10}.},
explicit examples of 10-dimensional solutions with a Lifshitz factor were constructed in \cite{BN10} and subsequently generalised in \cite{DG10,Donos:2010ax,Gregory:2010gx}, and, in the context of $\mathcal{N}=2$
gauged supergravity, in \cite{HPZ11,CF11}.

The four-dimensional Lifshitz solutions found in \cite{DG10} are supersymmetric and have scaling exponent $z=2$.  They can be obtained as circle reductions of Schr\"odinger solutions with $z=0$, which 
represent a plane wave propagating on the world-volume of D3-branes transverse to a Ricci-flat space and with non-zero RR and
NS magnetic 3-form flux. The examples found in \cite{DG10} can be thought of as a reduction of an $AdS_5$ type IIB vacua with the addition of some fluxes. 
A similar result has been proven in \cite{CF11}  at the level of gauge supergravity: any  ($d+1$)-dimensional gauge supergravity admitting 
$AdS_{d+1}$ vacua gives, upon reduction on a circle, a $d$-dimensional gauge supergravity admitting $Lif_d$ vacua with $z=2$. 
As we show below a similar result can  also  be obtained at the level of vacua in the 10-dimensional type II supergravities, avoiding all complications related to truncations and dimensional reductions.

In this note in fact  we analyse in detail the supersymmetry constraints on a large class of non-relativistic solutions which include and generalise the examples found in \cite{DG10}. 
We consider type II solutions corresponding to three-dimensional theories with anisotropic scaling in time $t$ and space $(x,y)$.  This class of solutions will include scale invariant theories,   domain walls interpolating between four-dimensional 
Lifshitz vacua  and more general solutions with only asymptotic, approximate scaling behaviour  or  hyperscaling violation. We will show that many supersymmetric four-dimensional Poincar\'e invariant backgrounds  in type IIB descend to analogous solutions with 
anisotropic scaling in $t$ and $(x,y)$. This result strengthens the correspondence between $AdS_5$ and $Lif_4$ vacua found in \cite{DG10,CF11}.  In particular, starting 
from supersymmetric domain walls  in type IIB interpolating between $AdS_5$ vacua, we often find solutions interpolating between the corresponding $Lif_4$ vacua. 

More generally we provide a  framework for constructing supersymmetric non-relativistic backgrounds
which can be useful also to study solutions with only asymptotic scaling and different IR behaviour,
corresponding for example to confining solutions. We hope to report on the subject in the next future. 
 

For convenience, in the rest of the Introduction, we summarise the relevant facts we need about supersymmetry.
In order to study non relativistic supersymmetric vacua we will use the formalism of Generalised Complex Geometry \cite{H02, G04}. In the case where the space is the (warped product)
\beq
\dd s^2 = e^{2 A} \dd s_4^2 + \dd s_6^2 
\eeq
of four-dimensional Minkowski or Anti de Sitter times an internal six-dimensional manifold $M_6$, 
the supersymmetry variations can be re-expressed as  a set of differential equations on the internal manifold.
More precisely, if $\eta^1_+$ and $\eta^2_+$ are two chiral spinors on $M_6$, we can define the bispinors
\beq
\label{6dps}
\Phi_\pm = \eta^1_+  \eta^{2\, \dagger}_\pm  \, , 
\eeq 
which, by Fierz identity, can be seen as sums of even, $\Phi_+$, and odd, $\Phi_-$,  forms on $M_6$\footnote{On $T(M_6) \oplus T^\ast(M_6)$, the polyforms  $\Phi_+$
and $\Phi_-$ correspond to pure spinors of positive and negative chirality, respectively.}. The supersymmetry variations can be shown to be equivalent to 
the following set of differential equations\footnote{Here  $\dd_H  = (\dd - H \wedge)$ 
and $\lambda (C_p) = (-1)^{[p/2]} C_p$ if $C_p$  is a form of degree $p$.}
\bea
\label{6dsusygen}
&& \dd_H (e^{2 A -\phi}  \Phi_1) =0 \, , \nn \\
&&  \dd_H ( e^{A -\phi}  \, {\rm Im} \Phi_2)  = 0  \, ,  \\ 
&&  \dd_H (e^{3 A  -\phi}   \, {\rm Re} \Phi_2)  = \frac{1}{8}\, e^{4 A}  \ast \lambda(F)  \, ,\nn
\eea
where  $F$ is the sum of the RR fluxes on the internal manifold, $H$ is the NS flux,  $\Phi_1 = \Phi_+$ and $\Phi_2 = \Phi_-$  
in IIA while for IIB  $\Phi_1 = \Phi_-$ and $\Phi_2 = \Phi_+$ \cite{GMPT}. 

A similar approach has been very recently used to study generic ten-dimensional vacua  \cite{T11}.
In this case, the  conditions for supersymmetry can be  reformulated in terms of intrinsic objects constructed with the type II supersymmetry parameters:  a  ten-dimensional  "spinor" $\Phi$
\beq 
\label{defspinor}
\Phi =  \epsilon_1 \bar{\epsilon}_2\, , 
\eeq
 and  two one-forms $K_1$ and $K_2$ 
\beq 
\label{defvectors}
K_{1 M} =  \frac{1}{32} \bar{\epsilon}_1 \Gamma_M \epsilon_1 \, , \qquad \,\,\,\, K_{2 M} =  \frac{1}{32} \bar{\epsilon}_2 \Gamma_M \epsilon_2  \, ,
\eeq
which annihilate  the supersymmetry parameters\footnote{$K_i\cdot \epsilon_i = K_{i\,M} \Gamma^M \epsilon_i$ is the Clifford multiplication.}
\beq
\label{vectors}
K_1\cdot \epsilon_1\, = \, K_2 \cdot \epsilon_2 \, =\, 0 \, .
\eeq
Each $K_i$ is null and gives rise to a basis of vielbeine $(e_{-i}\equiv K_i , e_{+i},e_I)$
which we normalize as
\beq
\label{norme}
e_{-i} \cdot e_{+i} = \frac{1}{2} \, .
\eeq

In this formalism the supersymmetry conditions are  \cite{T11}
\bea
\label{eq1}
&& \dd_H(e^{-\phi} \, \Phi) = - ( \tilde{K} \wedge  + \iota_K ) F \, , \\
\label{eq2}
&& \dd \tilde{K}  = \iota_K H \, , \\
\label{eq3}
&& ( e_{+ 1} \cdot \Phi \cdot e_{+2}, \Gamma^{MN} [\pm \dd_H(e^{- \phi} \Phi \cdot e_{+ 2} ) + e^\phi \dd^\dagger (e^{-2 \phi} e_{+2} ) \Phi - F ]) \, , \\
\label{eq4}
&& ( e_{+ 1} \cdot \Phi \cdot e_{+2}, [\dd_H(e^{- \phi} e_{+ 1} \cdot  \Phi ) -  e^\phi \dd^\dagger (e^{-2 \phi} e_{+2} ) \Phi -  F ]  \Gamma^{MN}  )  \, , 
\eea
plus  the  condition $L_K g =0$. The vectors $K$ and $\tilde{K}$ are 
\beq
\label{Ks}
K = \frac{K_1+ K_2}{2} \qquad \qquad \tilde{K} =  \frac{K_1- K_2}{2}  \, .
\eeq
In \eqref{eq3} and \eqref{eq4} we use the ten dimensional Mukai pairing  $(A,B) = (A \wedge \lambda(B))_{10}$ and, in \eqref{eq3}, the upper sign is for IIA and the lower is for IIB.

\vspace{0.3cm}

In this paper we will apply this formalism to the study of non-relativistic solutions of  Lifshitz type in type II supergravity. In Section 2 we discuss a class of non relativistic solutions of type IIA,
including and generalizing  \cite{DG10}, and we reduce the supersymmetry constraints to those of an auxiliary four-dimensional Poincar\'e invariant vacuum of type IIB plus a set of
algebraic and differential constraints on the fluxes. For reader convenience, we summarise our findings in Section \ref{4to3}. In Section 3 we  discuss the parallel case of non relativistic type IIB solutions. In Section 4 we discuss the particular case of $SU(3)$ structures
and we provide examples of $Lif_4$ solutions, domain walls and backgrounds with hyperscaling violation.


\section{Non-relativistic solutions in type IIA}
\label{LifIIA}

We look for static solutions in IIA supergravity  corresponding to three-dimensional theories with anisotropic scaling in time $t$ and space $(x,y)$.
We do not require to have exact scaling symmetry in order to accommodate also domain wall solutions interpolating between four-dimensional 
Lifshitz vacua  and more general solutions with only asymptotic or approximate scaling behaviour. As usual, one of the transverse directions will play the role of the radial
coordinate. 

On the basis of the existing examples of Lifshitz solutions \cite{DG10}  and of  the general structure of supersymmetric solutions in IIA \cite{T11}, we choose a metric of the form 
\beq
\label{IIAmet}
\dd s_{10}^2 =  - e^{2 A_1} \dd t^2 + e^{2 A_2} (\dd x^2 + \dd y^2) +  (e^1 )^2   + \dd s_6^2 \, ,
\eeq
with 
\beq
q e^1 =  \dd \varphi + \mu \, ,
\eeq
where $\varphi$ denotes a generic angular direction, $\mu$ is a connection on $M_6$ with curvature $\alpha \equiv \dd \mu$,  and $A_1, A_2$ and $q$ are  functions on $M_6$.  

According to this splitting we assume the
following structure for the fluxes\footnote{The RR fluxes are written in the democratic formulation \cite{Bergshoeff:2001pv}  as a formal sum $F=\sum_{k=0}^5 F_{(2k)}$,
with $k$ even in IIA and half-integer in IIB, 
subject to the condition $F= \ast_{10} \lambda(F)$.
We have  inherited from the democratic approach a somehow unnatural convention for the star product: $\ast C \wedge C = |C|^2 {\rm Vol}$.} 
\bea
\label{HIIA}
H^{IIA} &=&h + \dd(e^{01})  \, , \\
\label{FIIA}
F^{IIA}&=& - q (e^1 f + e^{0 x y}  \ast \lambda(f)  )
 + (1 + e^{0 1})  ( w  + e^{x y}   \ast \lambda(w)) \, ,
\eea
where we defined the three-dimensional vielbeine
\beq
e^0 =  e^{A_1} \dd t \, , \qquad \qquad e^x= e^{A_2} \dd x \,,  \qquad \qquad e^y= e^{A_2} \dd y \, , 
\eeq
and $f$ and $w$ denote formal sums  of odd and even  forms on the internal space $M_6$,
\bea
\label{fIIA}
 f &=&  f_1+ f_3+ f_5 \, ,\\
\label{wIIA}
w &=& w_0+w_2+w_4+w_6 \, . 
\eea
All star products are taken on the internal space  $M_6$.
For simplicity, we omit  the wedge products and define $e^{abc\cdots} = e^a \wedge e^b \wedge e^c \wedge \cdots$.  
Also,  for notational convenience, we write the type IIA dilaton $\phi_A$  in the form
\beq \label{dilatonIIA} 
e^{-\phi_{A}}= q \,  e^{-\phi} \, .
\eeq

The splitting of the metric is suggested by the conditions for supersymmetry. From  \eqref{eq1}-\eqref{eq4} it follows that 
 $K$ must be dual to  a Killing vector. For the static solutions we are considering we have a natural Killing vector corresponding to time translations. It is then natural to identify $K=e^0$.
The choice of metric (\ref{IIAmet})  corresponds to the general class of solutions where the second vector   identifies a special direction in the transverse space, $\tilde K=e^1$. 

The ansatz for the metric and fluxes makes it possible to reduce the search for vacua to a purely six-dimensional problem.
The equations of motion and the Bianchi identities for the fluxes reduce to a set of differential constraints for the forms $(\alpha,w, h,f)$ on $M_6$
and the supersymmetry conditions can be written in terms of the six-dimensional pure spinors \eqref{6dps}.  One of the main results of our analysis is that 
the six-dimensional data $(\phi,h,f,\Phi_\pm)$ must  satisfy the conditions for supersymmetry of a four-dimensional Poincar\'e invariant vacuum of type IIB,
 \eqref{6dsusygen}.
This will allow to construct non-relativistic 
solutions  with three space-time dimensions starting from known four-dimensional vacua. In the following sections we analyse the equations of motion and the supersymmetry constraints. We have summarised  the results in Section \ref{4to3}.

\subsection{Bianchi identities and equations of motion}
\label{Bianchi}

The fluxes have to satisfy the ten dimensional equations of motion and Bianchi identities
\bea
\label{RReom}
&&  \dd  F^{\rm IIA} - H^{\rm IIA} \wedge F^{\rm IIA}   = 0 \, , \\
\label{HBI}
&& \dd H^{IIA}  = 0 \, , \\
\label{Heom}
&& \dd (e^{- 2 \phi_{IIA}} \ast H^{IIA}) =  - \frac{1}{2} F^{IIA} \wedge \ast F^{IIA} \mid_8 \, .
\eea
The previous conditions reduce to the following constraints on the internal forms $h$ and $f$
\bea
&& \dd_h f  = 0 \, , \nonumber \\
&& \dd_h (q \, e^{A_1 + 2 A_2}  \ast \lambda(f)) = 0 \, , \nonumber \\
&& \dd h = 0 \, , \nonumber \\
&& \dd(q e^{A_1 + 2 A_2 - 2 \phi} \,  \ast h ) =   q \, e^{A_1 + 2 A_2} f   \ast f  \mid_4 \, ,
\label{fheom}
\eea
and for the forms $w$ and $\alpha$ 
\bea
&&  \dd_h w = \alpha f \, ,  \nonumber \\
&&  \dd_h (e^{2 A_2} \ast \lambda(w) ) =0 \, , \nonumber  \\  
&& \dd \,  \alpha =0 \, , \nonumber \\ 
&& \dd (  e^{2 A_2 -2 \phi} \, \ast \alpha ) = - e^{2 A_2} f  \ast w \mid_5\, , 
\label{waeom}
\eea
plus the the algebraic relation
\beq
\label{heomal}
e^{- 2 \phi} \alpha  \ast h  = w \ast f \mid_5 \, ,
\eeq
and the second order differential equation for $q$
\beq
\label{qeq}
 \ast   \dd (q e^{2 A_2 + A_1 -2 \phi}\,  \ast \dd (q e^{-A_1}))  =   e^{2 A_2} (  e^{- 2 \phi}|\alpha|^2 + |w|^2 ) \ .
\eeq

\subsection{Supersymmetry conditions}
\label{susy}

The supersymmetry conditions \eqref{eq1}-\eqref{eq4} can also be reduced to  a set of equations for the forms on $M_6$.
We will first  split the ten-dimensional spinors into the product of  a two-dimensional and an eight-dimensional spinor,  and then  further reduce the eight-dimensional spinors to 
six-dimensional ones. The form  \eqref{IIAmet} of the metric   suggests to choose $e^0$ and $e^1$ as preferred directions for our basis of  (real) ten-dimensional gamma matrices
\bea
\label{10gammas}
\Gamma^0 &=&  i \sigma_2  \otimes {\mathbb I}_{(8)}  \, , \nn \\
\Gamma^1 &=& \sigma_1 \otimes {\mathbb I}_{(8)}  \, , \\
\Gamma^I &=& \sigma_3  \otimes \hat{\gamma}^I \,  \qquad I= 1, \dots , 8 \, .   \nn 
\eea

In type IIA the ten dimensional spinors $\epsilon_1$ and $\epsilon_2$ are Majorana-Weyl and have opposite chirality.  Using the fact that 
$K=e^0$ and $\tilde{K}=e^1$ and the choice  \eqref{10gammas} of gamma matrices,  the conditions  \eqref{vectors}  become
\bea
\label{c1}
&& (\Gamma^0 + \Gamma^1) \epsilon_1 = 0 \, , \\
\label{c2}
&& (\Gamma^0 - \Gamma^1) \epsilon_2 = 0 \, , 
\eea
and imply  
\beq
\label{cc1}
\epsilon_1 = \begin{pmatrix} 1 \\ 0 \end{pmatrix}  \hat{\eta}_1  \qquad \qquad  \epsilon_2 = \begin{pmatrix} 0 \\ 1 \end{pmatrix}  \hat{\eta}_2 \, ,
\eeq
where the eight-dimensional spinors $\hat{\eta}_1$ and $\hat{\eta}_2$ are real and have both positive chirality\footnote{The ten-dimensional and the six-dimensional
chiral gammas are  
\beq
\Gamma_{10} = \Gamma^0 \ldots \Gamma^9 = \sigma_3 \otimes   \hat{\gamma}_9 \qquad \qquad  \gamma_7 = -i \gamma_1 \cdots  \gamma_6 \, . 
\eeq}. 
We further  split the eight transverse directions into two plus six. For the gamma matrices this gives
\bea
\hat{\gamma}^x &=& \sigma_1 \otimes {\mathbb I}  \, , \nn\\
\hat{\gamma}^y &=& \sigma_3 \otimes {\mathbb I}   \, , \\
\hat{\gamma}^m &=& \sigma_2 \otimes \gamma^m  \qquad m= 1, \dots , 6 \, . \nn
\eea
The spinors split accordingly 
\beq
\label{2+8split}
\hat{\eta}_ i = \sqrt{\frac{C}{2}} \Big[ \begin{pmatrix} 1 \\  i \end{pmatrix}  \eta^i_+  +
\begin{pmatrix} 1 \\  - i \end{pmatrix}  \eta^i_- \Big]  \qquad i=1,2 \, , 
\eeq
where $\eta^i_+$ are six-dimensional spinors of positive  chirality  and  $\eta_-= (\eta_+)^*$. 
It is convenient to redefine $\hat{\eta}_1\rightarrow \hat{\gamma}^{xy}  \hat{\eta}_1$ or, equivalently, $\eta_1\rightarrow - i \eta_1$. 

By construction  the vectors  $K_1$ and $K_2$   are given by
\bea
\label{Kivectors}
K_{(1) M} &= & \frac{1}{32} \bar{\epsilon}_1 \Gamma_M \epsilon_1 =  \frac{C}{16}  ||\eta^1_+ ||^2 (1, 1, 0,0,  \dots , 0) \, =    \frac{C}{16}  ||\eta^1_+ ||^2 (e^0 + e^1)_M  \, , \nonumber \\
K_{(2) M} &= &  \frac{1}{32} \bar{\epsilon}_2 \Gamma_M \epsilon_2 =  \frac{C}{16}  ||\eta^2_+ ||^2 (1, -1,0,0,   \dots , 0) \, =   \frac{C}{16}  ||\eta^2_+ ||^2  (e^0 - e^1)_M \, .
\eea
For simplicity, we restrict to the case where the norms of the six-dimensional spinors are equal:  $|| \eta_+^1 || = || \eta_+^2|| = || \eta_+|| $. We expect that, analogously to relativistic vacua,
this is the case of interest for backgrounds generated by D-branes. With equal norm for the spinors, the form of $K$ and $\tilde K$ is particularly simple,
\beq
\label{Kvectorspre}
K =   \frac{C}{16}  ||\eta_+ ||^2 \,  e^0  \qquad \qquad \tilde{K}  = \frac{C}{16}  ||\eta_+ ||^2 \, e^1 \, .
\eeq
We still need to require that $K$ is Killing. This fixes the functional dependence of $C$
\beq 
\label{CC}
C= 16  \frac{e^{A_1}}{ ||\eta_+ ||^2} \, ,
\eeq
and we finally have
\beq
\label{Kvectors}
K =   e^{A_1} \,  e^0  \qquad \qquad \tilde{K}  = e^{A_1}   \, e^1 \, .
\eeq
The normalization of $C$ is fixed in such a way that $K$ is precisely dual to $\frac{\partial}{\partial t}$.

\vspace{0.3cm}

The ten dimensional "pure-spinor"  also factorises into the product of  four and  six-dimensional  forms
\beq
\label{spinor10IIA}
\Phi =  \epsilon_1 \bar{\epsilon}_2 = - \frac{1}{2} (1 + e^{0 1}) \Phi_{(8)} \, ,
\eeq
with
\beq
\label{spinor8}
\Phi_{(8)}  =  \frac{16}{ ||\eta_+ ||^2}  \, e^{A_1}  \,  \{ {\rm Im}[(1 + i e^{x y})  \Phi_+] - {\rm Re}[(e^x - i e^y)  \Phi_-] \} \, . 
\eeq  
$\Phi_{\pm}$ are the six-dimensional pure spinors defined in \eqref{6dps}
\beq 
\Phi_+ = \eta^1_{+} \eta^{2 \, \dagger}_{+} \, , \qquad \,\,\,\,  \Phi_- = \eta^1_{+} \eta^{2 \, \dagger}_{-} \, .
\eeq
 
 \vspace{0.3cm}
 
Let us now consider the supersymmetry conditions \eqref{eq1}-\eqref{eq4}.
Equation  \eqref{eq2} is trivially satisfied. From equation  \eqref{eq1}  we obtain the six-dimensional equations
\bea
\label{6dsusy1}
&&  \dd_h(q \, e^{A_1-\phi} \frac{1}{ ||\eta_+ ||^2} \, {\rm Im} \Phi_+)  = 0  \, , \\
\label{6dsusy2}
&&  \dd_h(q\,   e^{A_1+2 A_2 -\phi} \frac{1}{ ||\eta_+ ||^2}  \, {\rm Re} \Phi_+)  = \frac{q}{8}\, e^{A_1 + 2 A_2}  \ast \lambda(f)  \, ,\\
\label{6dsusy3}
&& \dd_h(q\, e^{A_1+ A_2-\phi}  \frac{1}{ ||\eta_+ ||^2} \,  \Phi_-) =0 \, .
\eea
Notice that on the  right hand side  of equation  \eqref{eq1} $w$ disappears 
\beq
( \tilde{K} \wedge  + \iota_K ) F  = q\, e^{A_1} (1+ e^{01}) \wedge e^{xy}   \ast \lambda( f) \, .
\eeq

The conditions \eqref{eq3} and \eqref{eq4} can also be simplified. The computation is more involved and we give it  in  Appendix B.  What is important for our analysis is that 
they reduce to two independent sets of conditions, one for $f$ and one for $\alpha$ and   $w$.  All conditions for $f$ are 
 automatically satisfied once we fix 
 \beq
 \label{red}
 e^{A_1} = \frac{e^{2 A_2}}{q} \, ,  
 \eeq
while the conditions for $\alpha$ and $w$ are
\bea
\label{alphac1}
  {\rm Im}[  (\bar{\Phi}_+,   \frac{e^{ - \phi}}{   ||\eta_+ ||^2}  \alpha \cdot \Phi_+)_6 ] &=&  {\rm Im}[  (\bar{\Phi}_+, w)_6 ] = 0 \, ,  \\
\label{alphac1bis}
 {\rm Re}[ (\bar{\Phi}_+, ( \frac{ e^{ - \phi}}{   ||\eta_+ ||^2}  \alpha  \cdot \Phi_+ - i w))_6] &=& {\rm Re}[ (\bar{\Phi}_+,( \frac{ e^{ - \phi}}{   ||\eta_+ ||^2}    \Phi_+   \cdot \alpha+ i w) )_6]= 0 \, ,  \\ 
 \label{alphac2}
   {\rm Im}[ (\bar{\Phi}_+, \gamma^{m n} ( \frac{ e^{ - \phi}}{   ||\eta_+ ||^2}  \alpha  \cdot \Phi_+ - i w))_6] &=& {\rm Im}[ (\bar{\Phi}_+,( \frac{ e^{ - \phi}}{   ||\eta_+ ||^2}    \Phi_+   \cdot \alpha+ i w) \gamma^{m n}  )_6]= 0 \, ,  \\
 \label{alphac3}
( \Phi_-, \gamma^n (  \frac{4 \, e^{ - \phi}}{ ||\eta_+ ||^2}    \alpha \wedge \Phi_+ - i w))_6 &=&   ( \Phi_-,  (   \frac{4 \, e^{ - \phi}}{  ||\eta_+ ||^2}    \alpha \wedge \bar \Phi_+ - i w)\gamma^n )_6 = 0   \, ,
 \eea
where $\cdot$ denote the Clifford product and   $(A,B)_6 = (A \wedge \lambda(B))_{6}$ is the six-dimensional Mukai pairing.

With the redefinition (\ref{red}) the conditions (\ref{fheom}) and \eqref{6dsusy1}-\eqref{6dsusy3} 
involving $(f,h,\phi)$  become identical to the equations for the fluxes and  the supersymmetry 
conditions for  a type IIB vacuum with four-dimensional Poincar\'e invariance.  
This is one of the advantages of our formalism: the fields $(f,h,\phi)$ in our non-relativistic solutions are the same
as for four-dimensional relativistic solutions,  where many supersymmetric backgrounds are known. 
In the next  section we summarise our findings.

\subsection{Summary: from four to three dimensions}
\label{4to3}

Consider   a supersymmetric type IIB background
\beq
\label{I6dmet}
\dd s_{10}^2 =    e^{2 A} (\eta_{\mu \nu} \dd x^{\mu} \dd x^\nu )^2  + \dd s_6^2 \, ,  \qquad \mu=0,\cdots , 3 \, ,
\eeq
with dilaton $\phi$ and fluxes defined on $M_6$
\bea
\label{fluxes6d}
H_{IIB}  = h  \, , \qquad F_{IIB} =   f + e^{0 x y z}  \ast \lambda(f)    \, ,
\eea
and satisfying  the Bianchi identities and the equations of motion
\bea
&& \dd_h f  = 0 \, , \nonumber \\
&& \dd_h ( e^{4 A}  \ast \lambda f) = 0 \, , \nonumber \\
&& \dd h = 0 \, , \nonumber \\
&& \dd(  e^{4 A - 2 \phi} \,  \ast h ) =   e^{4 A} f   \ast f  \mid_4 \, .
\label{fheom0}
\eea

The type IIB supersymmetry parameters decompose as 
\beq
\epsilon_i = \zeta_+ \eta_+^i + \zeta_- \eta_-^i   \qquad i = 1,2 
\eeq
where $\zeta_+$ are four-dimensional Weyl spinors and $\eta_+^i$ are  six-dimensional spinors on $M_6$ normalized as 
\beq
\label{normal}
|| \eta_+^1 || = || \eta_+^2|| = \frac{c_+}{2} e^A  \, .
\eeq

The conditions for supersymmetry  can be written in terms of the six-dimensional pure spinors,  $\Phi_\pm$ of  \eqref{6dps},  as 
\bea
\label{6dsusy10}
&&  \dd_h(e^{A-\phi} \, {\rm Im} \Phi_+)  = 0  \, , \\
\label{6dsusy20}
&&  \dd_h(  e^{3 A -\phi}  \, {\rm Re} \Phi_+)  = \frac{c_+}{16}\, e^{4 A }  \ast \lambda  f  \, ,\\
\label{6dsusy30}
&& \dd_h( e^{2 A-\phi}  \,  \Phi_-) =0 \, .
\eea


\vspace{0.3cm}

Given such a four-dimensional supersymmetric  type IIB vacuum,
we can construct a non-relativistic supersymmetric solution in type IIA with metric (\ref{IIAmet}) and
\beq 
\label{map}
 e^{A_1}=\frac{e^{2 A}}{q}\, , \qquad e^{A_2}=e^{A}\, , \qquad e^{\phi_A}=\frac{e^\phi}{q} \, , 
\eeq
provided we find  on $M_6$ a two-form $\alpha$ and a polyform $w= \sum_{k=0}^3 w_{2k}$ satisfying the equations (coming from the 10-dimensional
Bianchi identities and equations of motion)
\bea
\label{waeomfin1}
&&  \dd_h w = \alpha f  \, ,   \\
\label{waeomfin2}
&&  \dd_h (e^{2 A} \ast \lambda(w) ) =0 \, ,   \\  
\label{waeomfin3}
&& \dd \,  \alpha =0 \, , \\ 
\label{waeomfin4}
&& \dd ( e^{2 A -2 \phi} \, \ast \alpha ) = - e^{2 A} f  \ast w \mid_5\, , \\ 
\label{waeomfin5}
&& e^{- 2 \phi} \alpha  \ast h  = w \ast f \mid_5 \, ,
\eea
and the supersymmetry constraints
\bea
\label{alphac1fin}
   {\rm Im}[   (\bar{\Phi}_+,   e^{ - \phi -A} \alpha \cdot \Phi_+)_6 ]  &=&    {\rm Im}[  (\bar{\Phi}_+, w)_6 ] = 0 \, ,  \\
\label{alphac1finbis}
 {\rm Re}[ (\bar{\Phi}_+,  ( e^{ - \phi -A}  \frac{2}{  c_+} \alpha  \cdot \Phi_+ - i w))_6] &=& {\rm Re}[ (\bar{\Phi}_+,( e^{ - \phi -A}  \frac{2}{  c_+}  \Phi_+   \cdot \alpha+ i w)  )_6]= 0 \, ,  \\
 \label{alphac2fin}
   {\rm Im}[ (\bar{\Phi}_+, \gamma^{m n} ( e^{ - \phi -A}  \frac{2}{  c_+} \alpha  \cdot \Phi_+ - i w))_6] &=& {\rm Im}[ (\bar{\Phi}_+,( e^{ - \phi -A}  \frac{2}{  c_+}  \Phi_+   \cdot \alpha+ i w) \gamma^{m n}  )_6]= 0 \, ,  \\
 \label{alphac3fin}
( \Phi_-, \gamma^n (  e^{ - \phi -A}  \frac{8}{  c_+}    \alpha \wedge \Phi_+ - i w))_6 &=&   ( \Phi_-,  (  e^{ - \phi -A}   \frac{8}{  c_+}   \alpha \wedge \bar\Phi_+ - i w)\gamma^n )_6 = 0   \, ,
 \eea
 where $\cdot$ denote the Clifford product and   $(A,B)_6 = (A \wedge \lambda B)_{6}$ is the six-dimensional Mukai pairing.
The function $q$ is determined by  the second order differential equation 
\beq
\label{qeqfin}
 \ast   \dd (e^{4 A -2 \phi}\,  \ast \dd (q^2 e^{-2 A}))  =  e^{2 A} (  e^{- 2 \phi}|\alpha|^2 + |w|^2 ) \ .
\eeq

\section{Non-relativistic solutions in type IIB}
\label{LifIIB}

The situation in type IIB is similar and we will be brief. Many solutions can be obtained by a plain T-duality. However there are a  few interesting features of the
solutions that deserve comments. We are interested in metrics of the form 
\beq
\label{IIBmet}
\dd s_{10}^2 =  - 2\,  q\,  e^{A_1} \dd t \,  \dd \varphi   + e^{2 A_2} (\dd x^2 + \dd y^2) +  q^2\, \dd \varphi^2   + \dd s_6^2 \, ,
\eeq
where, as before, $A_1, A_2$ and $q$ are  functions on $M_6$.  We introduce the null vielbeine
\beq 
e^- =  e^{A_1}(e^0 +e^1) \, , \qquad  e^+ = -\frac{1}{4} e^{-A_1}(e^0 -e^1) \, , 
\eeq
in terms of 
\beq
e^0 = e^{A_1} \dd t \, , \qquad  e^1= q\, \dd \varphi - e^{A_1} \dd t \, . 
\eeq
Notice that the Killing vector $\frac{\partial}{\partial t}$ is now null and dual to $e^- = q e^{A_1}\dd \varphi$.

\vspace{0.3cm}

The fluxes are
\bea
\label{HIIB}
H^{IIB} &=&h -  \frac{e^{-A_1}}{q}  e^- \alpha  \, ,  \\
\label{FIIB}
F^{IIB}&=& f + 2 e^{+ - x y}  \ast \lambda(f )
 + \frac{e^{-A_1}}{q}  e^-  ( w  + e^{x y}   \ast \lambda(w) ) \, ,
\eea
where  we introduced  formal sums $f$ and $w$ of odd and even  forms  defined on the internal space $M_6$ as in equations (\ref{fIIA}) and (\ref{wIIA}).
Again all star products are taken in $M_6$. 

\vspace{0.3cm}

Since the  ten-dimensional supersymmetry parameters in type IIB have the same chirality, we now have
\beq
\epsilon_1 = \begin{pmatrix} 1 \\ 0 \end{pmatrix}  \hat{\eta}_1  \qquad \qquad  \epsilon_2 = - \begin{pmatrix} 1 \\ 0 \end{pmatrix}  \hat{\eta}_2 \, ,
\eeq
where  the eight-dimensional spinors $\hat{\eta}_1$ and $\hat{\eta}_2$ are real and  both with positive chirality. 
The sign in $\epsilon_2$ is chosen for convenience.
We can still split the eight-dimensional spinors in two plus 
six as in equation (\ref{2+8split}). However now $K_1=K_2$, so that $\tilde K=0$ and $K$ is null
\beq 
K = e^- \, . 
\eeq 
Moreover, the ten-dimensional "pure-spinor" is now proportional to $e^-$
\beq
\label{spinor10IIB}
\Phi =  \epsilon_1 \bar{\epsilon}_2 = - \frac{1}{2} ( e^0 + e^1) \Phi_{(8)} \, .
\eeq

Exactly as before,  given a four-dimensional supersymmetric  type IIB vacuum corresponding to the pure spinors $\Phi_\pm$, with metric (\ref{I6dmet}), dilaton $\phi$ and fluxes $(h,f)$,
we can construct a non-relativistic supersymmetric solution with metric (\ref{IIBmet}) with
\beq  
\label{map2}
e^{A_1}=\frac{e^{2 A}}{q}\, , \qquad e^{A_2}=e^{A}\, , \qquad e^{\phi_B}=e^\phi \, , 
\eeq
provided we find  $M_6$  forms $\alpha, w$ satisfying the constraints \eqref{waeomfin1}-\eqref{qeqfin}.

\vspace{0.3cm}

There are a couple of interesting observations to be made. Since $K$ is null in type IIB, the equations of motion are not necessarily a consequence of the supersymmetry conditions.
As discussed in \cite{LT4}, all  Einstein and dilaton equations of motion follow from supersymmetry except perhaps the $(0,M\ne 0)$ components of the Einstein equations.
Indeed, in our case  the $(0,M\ne 0)$ components of the Einstein equations imply the conditions (\ref{heomal}) and (\ref{qeq}) which are not consequences of supersymmetry nor the Bianchi identities and equations of motion for the fluxes.

It is interesting to notice also that both sides of equation (\ref{eq1}) are proportional to $e^-$ and therefore to $\dd \varphi$. This allows to consider more general solutions where some of the functions and fluxes depend on $\varphi$. Examples of this type have been considered in \cite{BN10, DG10,  CH11, Balasubramanian:2011ua}. The generalization of our formalism to this case is straightforward. In this paper
we consider the simplest example of such construction, obtained by sending 
\beq 
\label{rot}
\eta_+^i \rightarrow e^{i k \varphi/2} \eta_+^i  \, . 
\eeq
Since $\Phi$ is proportional to $e^-$, equation (\ref{eq1}) is still satisfied. As discussed in  Appendix B, new terms appear instead in equations (\ref{eq3}) and (\ref{eq4}). As a result, in the  supersymmetry conditions  \eqref{alphac1}-\eqref{alphac2}  we must  replace 
\beq
\label{shift0}
\frac{\alpha}{4} \cdot \Phi_+ \rightarrow \frac{\alpha}{4} \cdot \Phi_+  + i\frac{k}{2}\Phi_+\, ,   \qquad    \Phi_+ \cdot \frac{\alpha}{4}  \rightarrow  \Phi_+ \cdot \frac{\alpha}{4}    -  i\frac{k}{2}\Phi_+\, .  
\eeq
We will discuss examples of this type  in Section \ref{SU(3)}.
 
\section{SU(3) structure solutions}
\label{SU(3)}

In the previous section we showed how, given a supersymmetric type IIB solution with four-dimensional Poincar\'e invariance  \eqref{I6dmet} and \eqref{fluxes6d}, one can
construct solutions, both in type IIA and type IIB supergravity, with a non-relativistic three-dimensional factor. In this section we will discuss some explicit examples of such
construction.  We will focus on a class of solutions where the internal manifold $M_6$ has SU(3) structure, since in this case the supersymmetry conditions for
the forms $\alpha$ and $w$ take a particularly simple form. We leave the study of  SU(2) structure solutions for a future work.

\vspace{0.3cm}

In rewriting the supersymmetry variations in terms of pure spinors on the internal manifold $M_6$, we make the assumption that the six-dimensional supersymmetry
parameters $\eta^1_+$ and $\eta^2_+$ are globally defined. Generically, two globally defined spinors reduce the structure group of $M_6$ to SU(2). However if 
they are parallel
\beq 
\eta^1 = \eta_+ \, , \qquad  \eta_+^2 = e^{i\theta} \eta_+ \, ,
\eeq
the structure group is SU(3).  In six dimensions any spinor $\eta_+$ is annihilated by three gamma matrices  
\beq 
(\gamma^{m} + i \gamma^{m + 3})\eta_+=0\, , \qquad  m=1,2,3 \, , 
\eeq
and defines a complex structure. The SU(3) structure can then be equivalently expressed in terms of  a $(1,1)$ (with respect to the complex structure)  two-form $J$ and a $(3,0)$ three-form $\Omega$, and the pure spinors \eqref{6dps} have the form
\beq 
\label{pureCY}
\Phi_+ = \frac{1}{8} e^{-i \theta} ||\eta_+ ||^2 e^{-i J} \, , \qquad \Phi_-=  -\frac{i}{8} e^{i \theta} ||\eta_+ ||^2 \Omega \, . \eeq
In ordinary four-dimensional vacua, the case $\theta=0$ corresponds to a manifold $M_6$ which is (conformally) Calabi-Yau.

\vspace{0.3cm}

For SU(3) structure manifolds, the  supersymmetry conditions for the forms $\alpha$ and $w$ can be simplified using the formulae discussed in  Appendix A,  in particular equation (\ref{Muk6}). With simple manipulations we can write the set of constraints \eqref{alphac1}-\eqref{alphac3} as
\bea
\label{alphac1su3}
 \eta_+^{\dagger} w \, \eta_+ &=&  \eta_+^{\dagger} \alpha \, \eta_+ = 0 \, ,  \\
 \label{alphac2su3}
   {\rm Re}[  e^{i \theta}   \eta_+^{\dagger} \gamma^{m n}  (  e^{-\phi - i \theta} \alpha - i w) \eta_+] &=& {\rm Re}[  e^{i \theta}   \eta_+^{\dagger}  (  e^{-\phi - i \theta}  \alpha + i w) \gamma^{m n}  \eta_+]= 0 \, ,  \\
 \label{alphac3su3}
 \eta_-^{\dagger} \gamma^{n}  (e^{-\phi - i \theta}  \alpha - i w) \eta_+&=&    \eta_-^{ \dagger} ( e^{-\phi + i \theta}  \alpha - i w)\gamma^{n}    \eta_+= 0   \, ,
 \eea
  or, equivalently,
 \bea
\label{alphac1su32}
 (\bar{\Phi}_+,     \alpha )_6 &=& (\bar{\Phi}_+, w)_6 = 0 \, ,  \\
 \label{alphac2su32}
   {\rm Im}[ (\bar{\Phi}_+, \gamma^{m n} ( e^{-\phi - i \theta}  \alpha - i w))_6] &=& {\rm Im}[ (\bar{\Phi}_+,( e^{-\phi - i \theta}  \alpha + i w) \gamma^{m n}  )_6]= 0 \, ,  \\
 \label{alphac3su32}
( \Phi_-, \gamma^n (   e^{-\phi - i \theta}  \alpha - i w))_6 &=&   ( \Phi_-,  (   e^{-\phi + i \theta}  \alpha - i w)\gamma^n )_6 = 0   \, .
 \eea


\vspace{0.3cm}

Before discussing explicit solutions, let us analyze the content of the supersymmetry 
conditions \eqref{alphac1su32}-\eqref{alphac3su32}.
Consider first solutions where the polyform $w$ only has a two-form component $w_2$. Then the general solution of the supersymmetry conditions  is obtained by requiring  that $\alpha$ and $w_2$ are $(1,1)$ and primitive\footnote{A two-form $C$ is primitive if its contraction with $J$ is zero, $C_{mn} J^{mn} =\eta_+^\dagger C \eta_+=0$; this is equivalent to $(\bar \Phi_+,C)=C\wedge J^2 =0$.}. The requirement that  
$\alpha$ and $w_2$ are  primitive is just equation (\ref{alphac1su3}), or, equivalently, (\ref{alphac1su32}). If $\alpha$ and $w_2$ are also $(1,1)$ all other equations are automatically  satisfied since any real two-form, $C_2$, $(1,1)$ and primitive satisfies
\beq
C_2\eta_+=\eta_\pm^\dagger C_2=0 \, .
\eeq 
In the special case $\theta=0$,  a $(2,0)$ component in $ ( e^{-\phi }   \alpha - i w_2)$ will also be  allowed. In fact, 
in this case 
\beq
 ( e^{-\phi }   \alpha - i w_2)\eta_+ =  \eta_-^\dagger ( e^{-\phi }   \alpha - i w_2) = \eta_+^\dagger ( e^{-\phi }   \alpha + i w_2) =0 \, .
 \eeq

In the general case, $\alpha$ is still required to be primitive. The conditions on the other forms are more involved.
Let us notice that  all conditions can be satisfied at once if 
\beq
\alpha\eta_+=w \eta_+= \lambda(w)\eta_+=0  \, , 
\eeq
since then $\eta_\pm \alpha=\eta_\pm w=0$\footnote{For a generic even polyform $A$, $A \eta_+ =0$ implies 
$\eta_+^\dagger \lambda(\bar{A}) = \eta_-^\dagger \lambda(A)=0$.}.

In type IIB, the modifications (\ref{shift0}) one has to make  to the supersymmetry conditions 
 when the spinor depends on  $\varphi$ 
greatly simplify  for $\theta=0$,  and become  equivalent to a shift in the zero form component of $w$: $w_0\rightarrow w_0- 2 k e^{-\phi}$.

\subsection{Examples}

Supersymmetric $Lif_4$ solutions based on $AdS_5$ vacua have been discussed in \cite{BN10, DG10, CF11,CH11}.  As we will discuss now,
solutions of this kind can be easily cast in our formalism.  In the same spirit,  we will show how  flows among $AdS_5$ vacua can give rise to flows among 
the corresponding $Lif_4$ vacua. 

\subsubsection{$Lif_4$ from $AdS_5$ solutions}
\label{AdS5}

A first example of solutions fitting the ansatz of Sections \ref{LifIIA} and \ref{LifIIB} is the family  found in \cite{DG10}, which we review here. 
They represent  type IIA and IIB  $Lif_4$ backgrounds where the internal manifold is a $U(1)$ fibration over a five-dimensional Sasaki-Einstein manifold $Y$.  
Both IIA and IIB solutions  are based on the standard type IIB vacuum of the form  $AdS_5\times Y$
\beq
\label{AdS5sol}
\dd s_{10}^2 =    r^2 ( \eta_{\mu \nu} \dd x^{\mu} \dd{x}^\nu)^2  +   \frac{1}{r^2}  (\dd r^2 +  r^2 \dd s_Y^2 ) \,  , 
\eeq
where  the internal manifold $M_6$  is a conformal Calabi-Yau cone over the Sasaki-Einstein $Y$.
The dilaton $\phi$ is constant and set to zero, and, in the  notations of Sections \ref{LifIIA} and \ref{LifIIB}, there is only a five-form  flux
\bea
\label{fluxesAdS5}
 h =0  \, , \qquad  \ast f  = 4 \frac{\dd r}{r}     \, .
\eea
The pure spinors are given by equation (\ref{pureCY}) with $\theta=0$ and  
\beq
J = \frac{1}{r^2} J_{\rm CY} \qquad \qquad \Omega = \frac{1}{r^2} \Omega_{\rm CY} \, .
\eeq
The supersymmetry conditions \eqref{6dsusy10}-\eqref{6dsusy30}  are trivially satisfied since $\dd J_{CY}=\dd \Omega_{CY}$=0. We normalized the spinors with $c_+=2$ (this choice implies $||\eta_+||^2= e^A$, cfr equation (\ref{normal})).
 
 \vspace{0.3cm}
 
Using (\ref{map}) we can write the associated non-relativistic type IIA solution \cite{DG10} as
\beq
\label{DGmetIIA}
\dd s^2_{10} = -\frac{r^4}{q^2} \dd t^2 + r^2 (\dd x^2 + \dd y^2) + \frac{\dd r^2}{r^2} + \frac{1}{q^2} (\dd \varphi +\mu)^2 + \dd s_Y^2 \, ,
\eeq
where $\dd \mu= \alpha$. The dilaton is 
\beq
e^{- 2 \phi} =  q^2 \, .
\eeq

The forms $\alpha$ and $w$ must satisfy  equations \eqref{waeomfin1}-\eqref{waeomfin5}. The simplest way to solve these constraints is to take only two-forms $\alpha$ and $w_2$ which are closed and co-closed on $Y$, or, equivalently harmonic on $Y$. These
exist on all Sasaki-Einstein manifolds with $b_2\ne 0$. They can be  primitive and of type $(1,1)$ so that, as discussed above, they preserve supersymmetry.  The fluxes then have the form \cite{DG10}
\bea
&& B =  \frac{r^2}{q^2}  \dd t \wedge (\dd \varphi +\mu) \, , \\
&& F_2 = w_2 \, , \\
&& F_4 = - 4 r^3 \dd t \wedge \dd x^1  \wedge \dd x^2  \wedge \dd r + \frac{r^2}{q^2}  \dd t \wedge (\dd \varphi +\mu) \wedge w_2\, .
\eea
The function $q$ on $Y$ must satisfy (\ref{qeqfin})
\beq 
\label{AdSY}
4 q^2 - \Box_Y q^2 = |\alpha|^2 +|w_2|^2 \, .
\eeq
Explicit solutions of this equation have been found in \cite{DG10} in the case of $T^{1,1}$  and some
$Y^{p,q}$. For $T^{1,1}$ the function $q$ is constant. Obviously we cannot find solutions of this type on $S^5$ since there are no harmonic forms on $S^5$.

\vspace{0.3cm}

It is interesting to look at  the form of the supersymmetry parameters. They satisfy conditions  (\ref{c1}) and (\ref{c2}).
For the solutions with $\eta^1_+=\eta^2_+$ we see from equation (\ref{cc1}) that
\beq
\label{c3}
\Gamma^{0xy}  \epsilon_1 =   \epsilon_2 \, ,
\eeq
where we took into account the redefinition $\hat{\eta}_1\rightarrow \hat{\gamma}^{xy}  \hat{\eta}_1$. This condition
is reminiscent of the original D3 brane condition. 

\vspace{0.3cm}

The case of type IIB is analogous. The metric is now of the form (\ref{IIBmet}) 
\beq
\label{DGmetIIB}
\dd s^2_{10} =  - 2\,  r^2\, \dd t \, \dd \varphi + r^2 (\dd x^2 + \dd y^2) + \frac{\dd r^2}{r^2} + q^2  \, \dd \varphi^2 + \dd s_Y^2 \, ,
\eeq
and the forms $\alpha$ and $w$ enter in the NS-NS and RR-RR fluxes as in (\ref{HIIB}) and (\ref{FIIB}). Notice that 
the fluxes are proportional to $\dd \varphi$ and the time-translation Killing vector is null. We still have solutions based on harmonic two-forms on $Y$ \cite{DG10}.

\vspace{0.3cm}

In type IIB we can also have solutions with an explicit dependence on $\varphi$ . These have been discussed in details in \cite{BN10, DG10,  CF11, CH11,Balasubramanian:2011ua}.  We just consider the simplest case. Set all the forms $\alpha$ and $w$ equal to zero, except for $w_0$. The resulting solution then can exist also on $S^5$. A non-zero $w_0$ corresponds to a one-form flux
\beq
\label{axion}
F_1 =  w_0\dd  \varphi  \, , 
\eeq
or, equivalently, to a linear profile for the axion $C_0=  w_0 \varphi $.
As discussed at the beginning of this section, with a $\varphi$ dependent spinor we can compensate the non-zero value of $w_0$ (since $\theta=0$) and have a supersymmetric solution. The equation (\ref{qeqfin}) for $q$ reads
\beq 
4 q^2 - \Box_Y q^2 =  |w_0|^2 \, , 
\eeq 
and can be solved with a constant function $q$. When $\varphi$ is compact, some quantization condition should be imposed on the parameters in order to have a consistent solution \cite{CH11}. The supersymmetric solution with non-compact $\varphi$
has been interpreted in \cite{Balasubramanian:2011ua} as describing a deformation of ${\cal N}=4$ SYM (or, more generally, of the four-dimensional superconformal theory associated with $Y$) in the presence of  a linear theta angle.
 
\subsubsection{Asymptotically $Lif_4$ solutions}

The ansatz described in Sections \ref{LifIIA} and \ref{LifIIB}  allows to show in an elegant way that many supersymmetric asymptotically $AdS_5$ backgrounds  in type IIB descend to analogous solutions with asymptotic $Lif_4$ vacua.
In particular, starting from  supersymmetric domain walls  in type IIB we expect to find  solutions interpolating between $Lif_4$ vacua.
In the gauge/gravity correspondence, a domain-wall connecting two $AdS_5$ vacua is interpreted as a RG flow between the corresponding $CFT_4$s \cite{GPPZ98, FGPW99}. 
The interpolation between $Lif_4$
vacua has obviously an analogous interpretation in terms of RG flows between $NRCFT_3$. 
\vspace{0.3cm}

As an example, we consider the case of 4d RG flows  with SU(3) structure based on internal manifolds that are  (non-conical) $CY_6$ 
\beq
\label{CY6}
\dd s_{10}^2 =   e^{2 A} ( \eta_{\mu \nu} \dd x^\mu \dd x^\nu)   +   e^{-2 A} \dd s_{CY_6}^2 \, ,   \qquad \mu=0,\, .. \, ,3  \, .
\eeq
The dilaton is constant and set to zero, and all fluxes are zero except for the five-form
\beq
\ast f = 4 \dd A \, ,
\eeq 
where $e^{-4 A}$ is required to be a harmonic function on $CY_6$. 
The pure spinors are given again by equation (\ref{pureCY}) with $\theta=0$, and 
\beq
J = e^{-2 A} J_{\rm CY} \qquad \qquad \Omega = e^{-2 A} \Omega_{\rm CY} \, .
\eeq

\vspace{0.3cm}

The first regular domain-wall which is conformally Calabi-Yau  was found in \cite{KM07} and interpolates between $AdS_5\times T^{1,1}$ and  
$AdS_5\times S^5$. It is obtained by displacing a large number of D3 branes on the resolved conifold. Near the position of the D3 branes a throat will emerge  recreating $AdS_5\times S^5$. The solution has a natural field theory interpretation in terms of baryonic VEVs. By moving in the baryonic moduli space of the conifold theory we can flow in the IR to pure $N=4$ SYM.

Similar solutions exist for all resolved $CY_6$  \cite{MS08}. In fact, it is well known that the web of four-dimensional quiver gauge theories associated to 
D3 branes sitting at 
conical Calabi-Yau singularities can be connected by RG flows induced by baryonic operators. On the geometrical side, the corresponding $CY_6$ are obtained by resolutions.
When the resolution of  the cone $C(Y_1)$ is only partial and the $CY_6$ has still local conical singularities of the form $C(Y_2)$ we can engineer a domain wall interpolating
between  $AdS_5\times Y_1$ and $AdS_5\times Y_2$ by putting D3 branes at the local singularity on the resolved Calabi-Yau.

\vspace{0.3cm}
From each of these solutions, we have type IIB domain-walls interpolating between the Lifshitz solutions corresponding to $Y_1$ and $Y_2$ 
\beq
\label{IIBmetdw}
\dd s_{10}^2 =  - 2\,    e^{2 A} \dd t \,  \dd \varphi   + e^{2 A} (\dd x^2 + \dd y^2) +  q^2\, \dd \varphi^2   +   e^{-2 A} \dd s_{CY_6}^2\, ,
\eeq
with  a linear axion 
\beq
\label{axion2}
F_1 =  w_0\dd  \varphi  \, . 
\eeq
The function $q$ is determined by
\beq 
\label{box}
-\Box_{CY_6} (q^2 e^{-2 A} ) = e^{-4 A} |w_0|^2 \, , 
\eeq
with $q$ becoming constant in the asymptotic regions. 

\vspace{0.3cm}

More generally, we  can have type IIA and type IIB solutions  with two-forms $\alpha$ and  $w_2$. The type IIB metric would be still of the form (\ref{IIBmetdw})  and in the type IIA  case we have
\beq
\label{IIAmetdw}
\dd s_{10}^2 =  - e^{4 A} \dd t^2 + e^{2 A} (\dd x^2 + \dd y^2) +  (e^1 )^2   +   e^{-2 A} \dd s_{CY_6}^2 \, .
\eeq
From equations \eqref{waeomfin1}-\eqref{waeomfin5} we see that  the two-forms $\alpha$ and $w_2$  must be  closed and co-closed on the $CY_6$
\beq 
\dd \alpha = \dd w_2 = 0 \, \qquad   \dd (\ast_{CY_6} \alpha) =  \dd (\ast_{CY_6} w_2) =0 \, .
\eeq
Such harmonic forms exist on resolved $CY_6$ with $b_2\ne 0$. Explicit solutions for the resolved conifold and its quotients can be found in \cite{LVP02, KMRW07}
and a general discussion in  \cite{MS08}. In this case $q$ is determined by
\beq 
-\Box_{CY_6} (q^2 e^{-2 A} ) = e^{-4 A}(|\alpha|^2 +  |w_2|^2) \, .
\eeq
In the original example \cite{KM07}  the solution of this equation becomes singular in the IR, since $S^5$ has no harmonic two-forms. For 
more general solutions interpolating between different Sasaki-Einstein manifolds we expect the existence of regular solutions.

The field theory realization of the $NRCFT_3$  dual to the existing $Lif_4$ vacua is still unclear. We may expect to understand them in terms of Chern-Simons gauge theories (see \cite{Balasubramanian:2011ua} for an attempt). 
The existence of a map between (many) $CFT_4$s to  $NRCFT_3$s and the corresponding flows suggests that it should be possible to understand and classify these $NRCFT_3$
in terms of the better known parent four-dimensional quiver gauge theories.  For this reason, it would be interesting to perform a full scan of the Sasaki-Einstein manifolds that give rise to $Lif_4$ vacua 
and of the allowed (regular) flows between them. Moreover, as recently pointed out \cite{Baggio:2011ha}, three-dimensional theories with $z=2$
have a trace anomaly which gives rise to a single central charge in models with a holographic $Lif_4$ dual. This central charge  is expected to decrease along a holographic flow on general grounds \cite{GPPZ98, FGPW99} and it would be very 
interesting to evaluate it for the models at hand.

\subsubsection{Solutions with hyperscaling violation}

The $Lif_4$ examples we have considered in Section \ref{AdS5} are scale-invariant;
the dilaton and all other scalar functions are required to be independent of the radial coordinate $r$. If we allow the dilaton, or other quantities, to
have a non-trivial profile in $r$ we can realize more general solutions with   a dynamical critical exponent $z$ and a hyperscaling violation exponent $\theta$ corresponding to the class of metrics \cite{Charmousis:2010zz,Dong:2012se}
\beq
\dd s^2 =  u^{-2(1-\frac \theta D)} \Big( -u^{-2 (z-1)} \dd t^2 + \sum_{i=1}^D  (\dd x^i)^2 + \dd u^2 \Big)  \, ,
\eeq
with $u=1/r$. The metric is conformal to the Lifschitz space-time but transforms as $\dd s^2 \rightarrow  \lambda^{2 \theta/D} \dd s^2$ under the rescaling 
\beq
\label{resc}
t \rightarrow \lambda^{ z} t \qquad \qquad x^i \rightarrow \lambda x^i  \qquad \qquad u \rightarrow \lambda u \, .
\eeq
Obviously these solutions have singularities for small or large $r$, and can be only considered as an effective description 
of the physics in some range of the radial coordinate.  For large and small $r$ we may expect the solution to have a 
different form, corresponding to an AdS or Lifshitz vacuum, or to a more general regular solution.  
As pointed out in \cite{Dong:2012se} a  very simple physical realization of such system with $z=1$ and 
$\theta=-\frac 1 3$  is given by D2 branes for a given range of values of $r$.

\vspace{0.3cm}

Starting with a general  IIB solution of the form (\ref{CY6}) with constant dilaton we can  obtain a non-relativistic solution with running dilaton  if  $q$ has an $r$-dependence (see (\ref{map}) and (\ref{map2})). The function $q$
satisfies equation (\ref{qeqfin}) which now reads
 \beq 
-\Box_{CY_6} (q^2 e^{-2 A} ) = e^{-4 A}(|\alpha|^2 +  |w|^2) \, .
\eeq
The general solution of this equation is obtained from a particular one by adding the solution of the homogeneous equation  $\Box_{CY_6} (q^2 e^{-2 A} )=0$. 
 
Let us consider, for instance, the case of conic Calabi-Yau manifolds of Section \ref{AdS5}, with $e^A = r$.
With zero internal forms $\alpha=w=0$ and $q= e^{-A}=1/r$, we obtain the non-relativistic type IIA solution
\beq
\label{nar}
\dd s_{10}^2 =  - r^6 \dd t^2 + r^2 (\dd x^2 + \dd y^2) +  r^2 d\varphi^2   +  \frac {\dd r^2} {r^2} + \dd s_Y^2 \, ,
\eeq
with $e^{\phi_A}= r$ and
\beq
H^{IIA} =  \dd( r^4 \dd t \wedge \dd \varphi) \, , \qquad\qquad  F_4 = - 4 r^3 \dd t \wedge \dd x^1  \wedge \dd x^2  \wedge \dd r \, . 
\eeq
There is also an analogous type IIB solution. Solutions of this kind, 
with $z=3$ and $\theta=2$ \footnote{The hyperscaling violation refers to the scaling of the effective $D+2=5$ dimensional metric in Einstein frame.}, 
have been discussed in \cite{Narayan:2012hk, Dey:2012tg}.  The solution only has a limited range of validity.
For example, for large $r$ the dilaton grows and invalidates the solution. We can have a different UV completion if we use 
the more general solution 
\beq
q^2 =  c_1 r^2 + c_2/r^2 \, .
\eeq
In this case  we obtain a metric that for large $r$ is a sort of  T-dual of the $AdS_5\times Y$ type IIB solution and  
reduces in the IR to the metric with hyperscaling violation. 

If we further add internal two- forms $\alpha$ and $w_2$  on $Y$ we can have more general solutions with 
\beq
q^2 = c_0 + c_1 r^2 + c_2/r^2 \, .
\eeq
where $c_0$ is a radial independent solution of (\ref{AdSY}).  For $c_2=0$ these solutions would interpolate between a T-dual of  $AdS_5\times Y$ in the UV and the $Lif_4$ solution 
discussed in Section \ref{AdS5} in the IR.


\section{Conclusions}

We have discussed a general framework to determine supersymmetric type II non-relativistic solutions with exact or asymptotic scale invariance. As already  emerged from previous investigations \cite{DG10,CF11},
there is a clear  correspondence between anisotropic $d$-dimensional vacua and relativistic solutions in $d+1$ dimensions. The known supersymmetric $Lif_4$ solutions have $z=2$ and descend
from $AdS_5$ vacua. This correspondence between $AdS_{d+1}$ and $Lif_d$ vacua is certainly intriguing and deserve further study. 
In particular it may be useful in  explicitly constructing  the  three-dimensional theories dual to Lifshitz vacua. 

Beside clarifying the correspondence between $d+1$ and $d$-dimensional vacua, our formalism can be applied to the search of new solutions.
In this paper we have only considered the simplest generalizations of the solutions found  in  \cite{DG10}, based on $CY_6$ four-dimensional vacua. 
There are other obvious directions of investigation.  For simplicity, we have only considered SU(3) structures. The case of  SU(2) structure is considerably more involved. 
However,  many interesting $AdS_5$ solutions with three-form fluxes, including the Pilch-Warner solution (PW) \cite{Pilch:2000ej} and the beta-deformation
of Sasaki-Einstein  backgrounds \cite{Lunin:2005jy}, have SU(2) structure   \cite{Minasian:2006hv, Butti:2007aq} and  we expect the existence of  corresponding $Lif_4$ solutions. 

 The formalism could be applied  also to the study of confining solutions. It would be quite interesting to see if  known relativistic four-dimensional confining solutions descend to non-relativistic solutions with 
 asymptotically Lifshitz scaling and a confining behaviour in the IR.  In the case of the obvious candidates (the Klebanov-Strassler \cite{Klebanov:2000hb}, the Maldacena-Nunez \cite{Maldacena:2000yy}
 and the interpolating baryonic branch  \cite{Butti:2004pk} solutions) it is not immediately obvious how to  find a set of polyforms $(\alpha, w)$ satisfying all constraints  \eqref{waeomfin1}-\eqref{waeomfin5} and \eqref{alphac1fin}-\eqref{alphac3fin} and 
 maintaining regularity. However, it is not excluded that a generalization of this construction will give interesting solutions.
We leave the detailed analysis of these and similar cases to  future work.

\vskip 1cm
\noindent {\bf Acknowledgements}
We wish to thank  A.~Tomasiello for interesting discussions.  M. P. is partially supported by the Institut de Physique Th\'eorique, du CEA. A.~Z.~is  partially supported by INFN and the MIUR-PRIN contract 2009-KHZKRX.


\begin{appendix}

\section{Notations and Useful Formulae}
\label{Mukai}

We use the notation of \cite{T11}  to which we refer for more details. To a differential form we can associate a bispinor via the Clifford map
\begin{equation}\label{eq:cliffordmap}
	C_k\equiv \frac1{k!} C_{M_1\ldots M_k} dx^{M_1}\wedge \ldots \wedge dx^{M_k} \ \longrightarrow \ \sla C_k \equiv \frac1{k!} C_{M_1\ldots M_k} \gamma^{M_1\ldots M_k}\ .
\end{equation}
The Clifford product is 
\begin{equation}\label{eq:gwedge}
	\gamma^M  C_k = (dx^M\wedge + \iota^M)  C_k
	\ ,\qquad
	C_k \gamma^M = (-)^k (dx^M \wedge - \iota^M) C_k \, ,
\end{equation}
where $\iota^M \equiv g^{MN}\iota_N \equiv g^{MN}\iota_{\del/\del x^N}$.  In this formulae and in the main text the "slash" symbol is usually understood. We keep 
only when needed to clarify the origin of few signs.  We take hermitian (imaginary and antisymmetric)  six-dimensional gamma matrices. The six-dimensional
chirality is defined through
\beq
\gamma_7 = -i \gamma^1\cdots \gamma^6 \, , 
\eeq
which on bispinors gives 
\beq 
\label{star} \gamma_7 \, \sla C = - i \ast \lambda ( \sla C) \, , 
\eeq 
where $\lambda (C_p) = (-1)^{[p/2]} C_p$,  if $C_p$  is a form of degree $p$.
This equation is consistent with our definition of the  star product 
\beq
\ast C \wedge C = |C|^2 {\rm vol}\, . 
\eeq

The six-dimensional Mukai pairing reads
\beq
\label{Muk6}
(A, B)_6 = (A \wedge \lambda B)_{6} = -   \frac{i}{8} (-1)^{{\rm deg} A}  {\rm Tr}( \gamma_7\,  (\sla {\bar A} )^\dagger \, \sla B)  \, .
\eeq
In using this equation, we must remember that the gamma matrices are  imaginary so that  one must be careful with  signs when converting the conjugate odd forms
into bispinor. The only relevant case for us comes with the complex conjugate of $\Phi_-$
\beq 
\slash{3}{25}{ (\bar \Phi_-)} = - \overline{(\slash{2}{15}{ \Phi_-})} \, .
 \eeq

The pure spinors 
\beq
\Phi_+ = \eta^1_+ \eta^{2 \dagger}_+\, ,   \qquad \qquad \Phi_- = \eta^1_+ \eta^{2 \dagger}_- \, ,
\eeq
satisfy the  identities
\beq
\label{pip0}
(\bar \Phi_+,\Phi_+ )_6 = (\bar \Phi_-, \Phi_-)_6 \, , \qquad (\Phi_+, Z \Phi_-)_6 = (\Phi_+, Z \bar \Phi_-)_6  = 0 \, ,  \qquad Z\in T  \oplus T^* \, ,
\eeq
Using (\ref{Muk6}) we can also derive the following identities
\bea
\label{Mukaipm}
&& ( \Phi_+, X \Phi_- Y)_6 =   -  (-1)^{\left[ \frac{{\rm deg} Y}{2}\right ]} (\Phi_-, X \Phi_+ Y)_6    = \frac{i}{8}  (\eta^{1 \dagger}_-  X \eta^1_+ ) (\eta^{2 \dagger}_- Y \eta^2_-) \, , \nn \\
&& (\bar{\Phi}_+, X  \Phi_+ Y )_6 =  (-1)^{\left[ \frac{{\rm deg} Y}{2}\right ]}  (\bar{\Phi}_-, X \Phi_- Y)_6  = -    \frac{i}{8} (\eta^{1 \dagger}_+  X \eta^1_+ ) (\eta^{2 \dagger}_+ Y \eta^2_+)
 \, ,   \\
&& (\Phi_+, X  \bar \Phi_- Y )_6 =  (-1)^{\left[ \frac{{\rm deg} X}{2}\right ]}  (\bar{\Phi}_-, X \Phi_+ Y)_6  = -    \frac{i}{8} (\eta^{1 \dagger}_-  X \eta^1_- ) (\eta^{2 \dagger}_+ Y \eta^2_-) 
\, , \nn
\eea
where $X$ and $Y$ are generic products of gamma matrices of the form $\gamma^{\{\mu_1\cdots\mu_k\}}$ acting in the Mukai pairing
via Clifford multiplication (\ref{eq:gwedge}). It can be useful to remember that $Y^T= (-1)^{\left[ \frac{{\rm deg} Y}{2}\right ]} Y$ and, as obvious from (\ref{pureCY}), $\eta^{i \dagger}_+  X \eta^i_+$
is non zero only for even $X$ and  $\eta^{i \dagger}_-  X \eta^i_+$ only for odd $X$. Other useful identities which follow from (\ref{star}) are
 \bea
 \label{star2}  
 &&(\bar\Phi_\pm, \gamma^m  \ast \lambda(f))_6= (\bar\Phi_\pm, \gamma^m (- i f))_6 \, , \nn \\
 && (\bar\Phi_\pm,   \ast \lambda( f)  \gamma^m)_6= (\bar\Phi_\pm,   i f \gamma^m )_6  \, .
 \eea

\section{The conditions for supersymmetry}

In this Appendix we show how to simplify conditions \eqref{eq3} and \eqref{eq4}. We will do it for the case of type IIA, type IIB being completely analogous.
We start from $K_1$ and $K_2$ given in (\ref{Kivectors}) 
\beq
\label{K12form}
K_{1} =   \frac{C}{16}  ||\eta_+ ||^2 (e^0 + e^1) \, ,  \qquad  \qquad  K_{2} =   \frac{C}{16}  ||\eta_+ ||^2  (e^0 - e^1) \, . 
\eeq
Setting $K_i = e_{-i}$,  we can define two basis of vielbeine as  $(e_{-i}, e_{+i},e_I)$, with 
\beq
\label{eplus}
e_{+1} = \frac{1}{4} e^{- A_1} ( e^1 - e^0)  \qquad \qquad e_{+2} = - \frac{1}{4} e^{-A_1} (e^0 + e^1) \, .
\eeq
where we use the normalization of  (\ref{norme}).

\vspace{0.3cm}

Using an identity analogous to (\ref{Muk6}) (see also  formula (B.33) in \cite{T11})
\beq
\label{ident}
( e_{+1}\cdot  \Phi \cdot e_{+2} , C ) = \frac{1}{32} (-1)^{{\rm deg} \Phi}  \bar{\epsilon}_1 e_{+1} C e_{+2} \epsilon_2 \, ,
\eeq
where $C$ is a generic bispinor, it is easy to verify that
\beq
\label{id1}
( e_{+1}\cdot  \Phi \cdot e_{+2} , C\cdot e_{+2} ) =  ( e_{+1}\cdot  \Phi \cdot e_{+2} , e_{+1} \cdot C )  =0
\eeq
and
\beq 
\label{id2}
( e_{+1}\cdot  \Phi \cdot e_{+2} , \Gamma_{10} C ) = ( e_{+1}\cdot  \Phi \cdot e_{+2} ,  C )  \, . 
\eeq
The first identity follows from the vanishing of the square of the null vector $e_{+i}$ and the second from the fact that $\epsilon_1$ has positive chirality.

\vspace{0.3cm}

Let us now consider  \eqref{eq3}.  The left-hand side of the Mukai paring gives 
\beq
e_{+1} \cdot \Phi \cdot e_{+2} = \frac{1}{8} C  e^{- 2 A_1} (1 - e^{01}) \Phi_8 \, .
\eeq
The $\dd_H$ terms can be written as 
\bea
\label{de2}
 \dd_H (e^{-\phi_A} \Phi \cdot e_{+2})  &=&  \dd_H (e^{-\phi_A- A_1} \frac{C}{4} \Phi_8 (e^0+e^1)) \nn \\
 &=&  \dd_h (e^{-\phi_A- A_1} \frac{C}{4} \Phi_8) \cdot  (e^0+e^1)   +  e^{-\phi_A- A_1} \frac{C}{4} \Phi_8 \dd (e^0+e^1) \nn \\
 && -\dd (e^{01}) \wedge e^{-\phi_A- A_1} \frac{C}{4} \Phi_8 (e^0+e^1)   \nonumber \\
&= &  \dd_h (e^{-\phi_A- A_1} \frac{C}{4} \Phi_8) \cdot  (e^0+e^1) \nn \\
&& + e^{-\phi_A- A_1} \frac{C}{4} \Phi_8 ( \dd A_1 e^0 - \frac{\dd q}{q} e^1 + \frac{\alpha}{q} (1+e^{01})) \, ,
\eea
while,  from \eqref{eplus}  it follows  immediately that
\beq
 \dd^\dagger (e^{-2 \phi_A} e_{+2} ) = 0 \, .
\eeq
Finally, the fluxes can be written as
\beq
\label{ffl}
F^{IIA} =  (1 + \Gamma_{10} )[-  q \, e^1 f +  (1 + e^{01}) w]  \, ,
\eeq
and, when inserted in \eqref{eq3}, we can use \eqref{id2} to replace $\Gamma_{10} $ in (\ref{ffl})  with the identity matrix.  
Notice also that the first term on the right hand side of equation (\ref{de2}) does not contribute to \eqref{eq3} because of the identities \eqref{id1}. Similarly, 
we can also manipulate the remaining terms dropping all pieces  of the form $( \, )\cdot (e^0+e^1)$. We finally obtain 
\beq
\label{M10}
 ((1-e^{01})  \Phi_8, \Gamma^{MN} [ ( e^{-\phi_A- A_1} \frac{C}{8} (\dd A_1+\frac{\dd q}{q})\Phi_8  +  q f) (e^0-e^1)    + (e^{-\phi_A- A_1} \frac{C}{4\, q }  \alpha \, \Phi_8 - 2 w) (1+e^{01})]) \, .
\eeq

The ten-dimensional Mukai pairing (\ref{M10}) can be reduced to a six-dimensional one using the explicit form of $\Phi_8$ given in equation (\ref{spinor8}). We need to distinguish various cases for $M$ and $N$.

When $M=0$ or $M=5$ the only non-zero contributions come from the term  proportional to $e^0-e^1$ on the right hand side.  Not unexpectedly,  these contributions are similar to those found in the analysis of four-dimensional vacua in  \cite{T11}. 
When $N=x,y$  most of the terms vanish due to (\ref{pip0}) and we obtain the constraint
\beq
\label{fconst1}
(\Phi_-,  f)_6 = 0 \, ,
\eeq
while $N=n$ gives 
\beq
\label{fconst2}
 {\rm Im}[( \bar \Phi_+, \gamma^n [ e^{-A_1 - \phi_A} \frac{C}{4} (\dd A_1+ \frac{\dd q}{q}) \Phi_+  - q \,  \ast \lambda f])] =0 \, ,
 \eeq
 where we  used (\ref{Mukaipm}) and (\ref{star2}). These two conditions are actually implies by the supersymmetry equations. Indeed, using  \eqref{star2} and   \eqref{6dsusy2},  
 \eqref{fconst1}  can be written as
  \beq
 \label{pip1}
(\Phi_-,  \ast  f)_6 \sim ( \Phi_-, \dd_h  \Phi_+)_6 = (\dd_h  \Phi_-,  \Phi_+)_6 =0 \, 
 \eeq 
due to  \eqref{6dsusy3} and (\ref{pip0}). Since \eqref{6dsusy1} and \eqref{6dsusy2} can be combined in 
\beq
\dd_h(\frac{C}{2} e^{A_2 - \phi_A} \bar \Phi_+) = -  \frac{C}{2} e^{A_2 - \phi_A} \dd A_2 \Phi_+  +  q  e^{A_1+ A_2} \ast \lambda f  \, ,
\eeq
we see that also  \eqref{fconst2} can be written as 
\beq
\label{pip2}
{\rm Im} [( \bar \Phi_+, \gamma^n \dd_h(\frac{C}{2} e^{A_2 - \phi_A} \bar \Phi_+))_6 =0 \, ,
\eeq
provided we set $e^{A_1} =e^{ 2 A_2}/q$. This equation is automatically satisfied since $\Phi_+$ is a  pure spinor. Indeed,  
\beq
(\bar \Phi_+, X\bar \Phi_+)
\eeq
can be different from zero only for an insertion $X$ of six gamma matrices.  On the contrary  $d_h$ can bring a maximum of three gamma since it is odd and $\dd$ 
can only change the complex type $(p,q)$ of a form by a maximum of two units in $p$ and $q$.

For  all other values of $M$ and $N$,  (\ref{M10}) gets contributions only from the term  proportional to $(1 + e^{01})$.   $M=0$ and $N=5$ and $M=x,$ $N=y$ give 
\beq
\label{alphac1A}
 (\bar{\Phi}_+,  \frac{C}{16 \,  q} e^{-A_1 - \phi_A}  \alpha \cdot \Phi_+ - i w)_6 = 0 \, ,
 \eeq
where $\cdot$ denote the Clifford product, while  for $M=m$ and $N=n$ we obtain
\beq
\label{alphac2A}
 {\rm Im}[ (\bar{\Phi}_+, \gamma^{m n} (\frac{C}{16 \,  q} e^{-A_1 - \phi_A}  \alpha \cdot \Phi_+ - i w))_6 ] = 0 \, .
\eeq
Finally  $M=x,y,$ and $N=n$ give 
\beq
\label{alphac3A}
( \Phi_-, \gamma^n (  \frac{ C}{4 \, q}  e^{-A_1 - \phi_A}  \alpha \wedge \Phi_+ - i w))_6 = 0   \, .
\eeq
In deriving the above conditions, we used  repeatedly the identities (\ref{Mukaipm}). Using (\ref{CC}) we recover half of the conditions \eqref{alphac1}-\eqref{alphac3}.
Equation  \eqref{eq4} can be treated in a similar way and gives the remaining half of the conditions.

\vspace{0.3cm}

The computation in type IIB  is  similar and it will not be reported here.  Let us simply note that, with spinors depending on $\varphi$ as in (\ref{rot}),
there is an extra contribution from $ \dd_H (e^{-\phi_A} \Phi \cdot e_{+2})$
coming from the component of $\dd \Phi_8$ along $e^-$; this term is responsible for  the replacement (\ref{shift0}).

\end{appendix}

\providecommand{\href}[2]{#2}\begingroup\raggedright\endgroup

\end{document}